# Rapid, Impartial and Comprehensive (RIC) publishing:
# A new concept for scientific journals

Sergey I. Bozhevolnyi

*Institute of Technology and Innovation (ITI), University of Southern Denmark, Odense M*

*seib@iti.sdu.dk*

**Abstract:** Publishing scientific journals governed by editors relying on anonymous peer reviewing is slow (even one round of reviewing involves several communications between authors, editor and reviewers), partial (arguments of authors can hardly overrule those of reviewers) and not using all available scientific material (even the most thorough and insightful reviews remain for the eyes of authors and editors only). Here I propose a new concept for scientific journals that ensures rapid, impartial and comprehensive (RIC) publishing. RIC concept is based on implementation of two novel publishing principles: the first (rapid) editorial screening of a submitted manuscript should result in its either "rejection" or "acceptance with optional revisions", and, in the latter case, the optionally revised (taking into account open reviews) paper should be published along with all (positive and negative) reviews, presenting thereby to the scientific community all available scientific material on the topic in question.

Any research should inevitably result in a publication not only because of the infamous principle "publish or perish" [1] but also because doing research is about facts that are ascertained and documented. Hence publications that establish intellectual property rights and provide the means of assessing the conducted research. One of the key components of publishing is the peer review process, which is implemented in one or another way in all scientific journals and which should secure the quality of publications. It is however widely accepted that the traditional peer review system has a number of disadvantages [2-4]. Among many approaches to the improvement of traditional system that are currently being considered [3], the arguably main approaches are masking the identity of both authors and reviewers (double-blind) and public peer review [4]. The fact is that double-blinding is very difficult to accomplish since authors would almost inexorably reveal themselves by choice of the topic, instrumentation and/or theoretical models used, not to speak about self-citation. Public peer review seems more promising but the described so far various implementations, very often involving two-stage reviewing [3,4], are rather bulky and complex and might contribute to frictions in the scientific community, when the authors would believe themselves being maltreated by the reviewers [4]. In this communication, after briefly considering the conventional peer review system, I suggest an original way of combining open review and fair treatment of both authors and reviewers, coming up with a new concept of publishing that is free of above drawbacks.

The process of getting a paper published is, as so well known to any scientist, "a long and windy road" that takes time and efforts out of the only source available – research. Indeed, even in the simplest case of one round of reviewing with minor revisions required (Fig. 1), a submitted paper is dormant (~ 1 week) while an editor identifies reviewers, after which the reviewers assemble their views (2-4 weeks, often with delays), which are subsequently assessed by the editor (~ 1 week), who then conveys the decision to authors. Note that the latter are most eagerly awaiting the decision all this time, since it would strongly influence the next step in their research (e.g., new experiments might be needed, hence the setup used should not be reassembled). The situation becomes worse if (as often is the case) major revisions are required, and reviewers asked to see the manuscript again to decide whether those were sufficiently close to the required ones (Fig. 1). Finally, the whole business might take a nightmarish direction with the second round of revisions required, which is still very much in practice despite the usual "firm journal policy to allow only one revision".

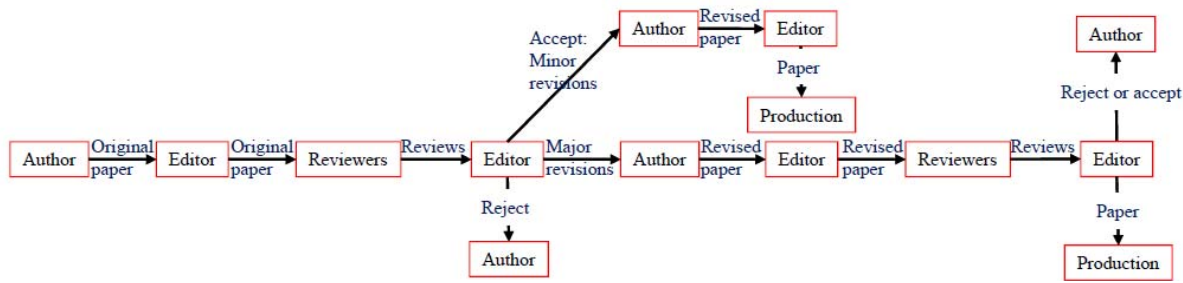

Fig. 1. Schematic of usual manuscript treatment on the way from authors to paper production.

All this hard work should be evaluated from the appropriate perspective: <u>majority</u> of the papers prepared for publication are getting published, although not always in the journals of their original submission. Moreover, most of the finally published papers are <u>marginally</u> different from those originally submitted: a typical list of changes to a manuscript takes less than one page while a paper is many pages long. Society would certainly be much better served if all energy of talented and knowledgeable scientists, authors and reviewers, used in this way would be directed to what they are so good at doing – research.

I am sure that all authors were subjected or, at least, thought that they were subjected to partial (unfair) paper treatment quite a few times. In my opinion, such an unfortunate treatment is an inevitable possibility in the current publishing process based on anonymous reviews. "Inevitable possibility" is not an oxymoron – it means exactly that: unfair treatment inevitably happens, albeit not always. Reviewing is by definition subjective and, being perfectly aware of having an upper hand in arguments with authors (while remaining anonymous to those), it is so easy to casually remark that "it is a solid paper but I feel that the advance achieved is incremental..." or "I am not really convinced that this paper deserves to be published in this journal..." – there is typically no backing up of this kind of statements with meaningful explanations and references. It is equally easy to ask conducting additional experiments or simulations that in fact surmount the reported ones. Also requirements of citing several not really relevant papers that have common authorship are becoming more of a rule than an exception. It is curiously to note that virtually all scientists are both authors and reviewers, just at different times and with respect to different papers. Having the experience in both roles might indeed be destructive for the whole publishing process, since reviewers are well aware (all too well!) of what can be required from authors to stall or even effectively block the paper. The above insinuations sound rather ugly and might be a bit overdramatic – research is not exactly a cutthroat business, but anyway… it surely is highly competitive! To conclude: in the publishing system, which is based on anonymous reviewers whose recommendations are seldom overruled by editors, it is practically impossible to ensure impartial consideration of submitted papers.

Besides the aforementioned problematic aspects of publishing, there is another one that is very seldom discussed: withholding an important part of accumulated scientific material – reviews. Indeed, it is such a pity to keep well prepared and elegantly formulated reviews only for the eyes of authors and editors. The point is that reviewers can and do write excellent reviews (though not always) backed up with meticulous and deeply thought through arguments, which can even be accompanied with simulations and figures, giving new explanations and profound insights for the considered phenomena. At the same time, from the point of view of a reviewer, it might indeed seem a waste of time writing a profound and well formulated review for the benefit of paper authors only. Such an understandable viewpoint has definitely negative implications to the aforementioned problem of partial paper treatment in this competitive scientific world. More importantly, keeping this kind of reviews undisclosed does certainly seem a waste of knowledge for the scientific community. Overall, judging from the prospective of having decent reviews, one would invariably conclude that the current publishing procedure is far from being comprehensive, withholding a great deal of pertinent scientific material from the public domain.

In order to avoid the publication process being slow, partial and incomplete, I suggest a novel and cardinally different approach ensuring rapid, impartial and comprehensive (RIC) publishing. The essence of RIC concept is rather simple and forthright: the first editorial screening should promptly select manuscripts for further processing with the formulation "accepted with optional revisions", informing immediately their authors along with sending the manuscripts to reviewers. The completed reviews should be sent to both editors and authors, so that the latter can exercise their right of optionally revisions, and, following this revision, the paper in final redaction is to be published along with all (positive and negative) reviews. The latter should be duly identified with DOI and other credentials (author, title, pages etc.) in the same way as papers.

Thus implemented publishing process would be very compact (Fig. 2) and fast reducing the decision waiting time to an absolute minimum, a feature that is of paramount importance for all authors. Nobody writes a paper for it to be rejected, but if it is to be that way then the sooner the better. There is no question: in the course of reviewing preceding the paper rejection, authors do receive serious criticism and comments that are useful and often indispensable for straightening and improving the paper. But the fact is that a paper becomes rejected when it cannot be improved and made suitable for the publication, at least from the point of view of reviewers. Assuming that most of the papers do bear a rationale worth publishing, one logically concludes that either the choice of journal was erroneous or the editorial choice of reviewers was unfortunate. In either case, the sooner authors would start looking for another journal the better!

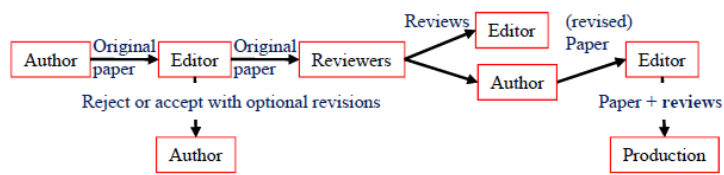

Fig. 2. Schematic of the manuscript treatment on the way to paper production when using RIC concept.

It should be pointed out that open reviewing alone would not suffice. It is true that open reviewing would inevitably be (much) better balanced, however very few reviewers would agree to open their names (the latter fact is well established via various surveys). Why is that? In my opinion, the main reason is that the open review process implemented on its own (i.e., without other changes) would in a way reverse the situation placing a much heavier burden on reviewers, whom it would become even more difficult (if possible at all) to find. Such a turn is caused not only by the issue of additional time one would have to spend on more careful phrasing and substantiated criticism but also by the problem of keeping agreeable and collaborative relationships with colleagues. In contrast, RIC concept would solve these problems by ensuring symmetry in the positions of authors and reviewers, securing thereby impartial paper treatment. Indeed, by publishing all (screened) papers irrespectively of reviews that would be published as well, the reviews would cease to be decisive for the paper acceptance and become more like detailed comments published on equal footing with papers, while keeping the original role of helping authors to improve their papers before the publication.

In addition, the implementation of RIC concept would make the reviewing job more attractive by removing the heavy burden of responsibility for the decision on publication (also removing seduction of stalling the paper) and offering a direct way of publishing reviewer's opinion on the paper topic. The latter is just as important for reviewers as getting a paper published for authors: it does happen that a negative review is overruled by an editor because of other positive reviews, wasting the time spent on preparing such a review and depriving the scientific community from hearing "the second opinion". Overall, the material published according to RIC concept (paper and reviews) would inevitably be richer in the scientific content than the paper itself, even after its very careful revision.

It is easy to see that all the RIC components are to some extent present in currently published journals. The first editorial screening is widely being used, becoming more of a rule than exception. Invited review papers are, in practice, always published after implementing minor revisions required by the reviewers – this is equivalent to being immediately accepted with optional revisions as integrated in the RIC concept. The procedure of expert comments (positive and solicited by editors) being published simultaneously with original papers is implemented, for example, in *Nature* journals (section "News and Views").[1] At the same time, negative comments offered voluntarily to previously published papers can be found in *Physical Review Letters*.[2] In both cases, these comments are often referred to and sometimes even independently on the original paper discussed in the comment. It should be noted that both open reviews and publication of reviews have been discussed in questionnaires sent out by various journals. I think that it is the combination of open <u>published</u> reviews and immediate editorial decision on papers that makes RIC concept being really balanced and foolproof.

---

[1] Quite often, the authors of these comments are chosen among reviewers of the corresponding papers.

The role of editorial screening would become very important in RIC publishing, and one can suggest several approaches for composing an efficient and suitably qualified editorial board, e.g., by identifying the most active scientists through publications and conferences and periodically renewing the editorial board. Nevertheless, it is clear that there is no way one can guarantee the screening process to be impeccably fair, but getting a verdict very fast would really minimize the potential damage. Just imagine waiting during weeks for reviews that turn out being not only negative but also quite useless in its content.

I would also venture to argue that full implementation of RIC concept would place the strictest publication selection in the hands of authors rather than editors. It would indeed be very risky for the paper authors themselves to submit a dubious paper, which could be accepted and published along with reviews demonstrating that the published results are wrong or misleading or simply nonsense.[3] At the same time, reviewers would realize that their responsibility of analyzing a paper under review would increase, since their endorsement or criticism would become public. The main responsibility of editorial screening would therefore be rather in selecting suitable (by topic, general or specific interest, etc.) than faultless papers.

Bottom line is that, in my opinion, authors and reviewers that are already doing great and conscientious work would probably not see any difference when switching to the suggested system, since their self-imposed requirements are strong enough. At the same time, getting paper published would really be straightforward and fast, and, on the top of that, the reviewing would also result in publications, becoming useful to many readers and not only to the paper authors. The most important advantage of the suggested publication procedure is in saving huge amount of time and many efforts (cf. Figs. 1 and 2), while ensuring that all scientific material accumulated by authors and reviewers on the topic in question becomes available for the whole scientific community. In addition, RIC publishing would invigorate open scientific discussions, at least between authors and reviewers, resulting eventually in establishment of higher ethical standards of personal communications and cleaner moral climate in the scientific community. Finally, there are considerable advantages of RIC concept from the viewpoint of publishers as well, such as simplification and acceleration of the publishing process, not to speak of getting the reputation of fair publishing. Overall, I think that there is a good chance for RIC concept to be found suitable for many (existing and coming up) scientific journals, whose implementation of RIC concept would greatly contribute to their popularity among scientists.

---

[2] These comments could be more interesting than the papers commented upon, as for example noticed during the debate on the issue of "perfect lensing".

[3] In this case, the authors would most probably revise their paper as much as possible saving the paper (if there was something to save, which would be a positive outcome) or facing the music!

Acknowledgement: I am very grateful to Elena Bozhevolnaya for forcing me into thinking along these lines by saying: "Stop complaining about wasting your time on reviewing and answering to reviews and do something about it!" as well as for numerous discussions we had on this subject. I acknowledge also the community of Château d'Azans (Dole, France) for creating a friendly atmosphere facilitating my work on this paper.